\documentclass{aa}
\usepackage{color}
\usepackage{natbib}
\usepackage{amsmath}
\usepackage{graphicx}
\usepackage{txfonts}
\begin{document}

   \title{Migrating Jupiter up to the habitable zone: \\ 
         Earth-like planet formation and water delivery}

   \author{ L. A. Darriba \inst{1,2}
           \thanks{ldarriba@fcaglp.unlp.edu.ar},
           G. C. de El\'ia \inst{1,2},
           O. M. Guilera \inst{1,2}
          \and
           A. Brunini \inst{3}
%         \thanks{}
          }

   \offprints{L. A. Darriba
    }

  \institute{Instituto de Astrof\'{\i}sica de La Plata, CCT La Plata-CONICET-UNLP \\
   Paseo del Bosque S/N (1900), La Plata, Argentina.
   \and Facultad de Ciencias Astron\'omicas y Geof\'\i sicas, Universidad Nacional de La Plata \\
Paseo del Bosque S/N (1900), La Plata, Argentina.
   \and Universidad Nacional de La Patagonia Austral. Unidad Acad\'emica Caleta Olivia \\
   Ruta 3 Acceso Norte, Caleta Olivia (9311), Santa Cruz. CONICET
                }

   \date{Received / Accepted}

%-------------------------------------------------------

\abstract
{Several observational works have shown the existence of Jupiter-mass planets covering a wide range of semi-major axes around
Sun-like stars.
}
{We aim to analyse the planetary formation processes around Sun-like stars that host a Jupiter-mass planet at
intermediate distances ranging from $\sim$1 au to 2 au. Our study focusses on the formation and evolution of terrestrial-like
planets and water delivery in the habitable zone (HZ) of the system. Our goal is also to analyse the long-term dynamical
stability of the resulting systems.
}
{A semi-analytic model was used to define the properties of a protoplanetary disk that produces a Jupiter-mass planet around the snow line, which is located at $\sim$2.7 au for a solar-mass star. Then, it was used to describe the evolution of embryos and planetesimals during the gaseous phase up to the formation of the Jupiter-mass planet, and we used the results as the initial conditions to carry out N-body simulations of planetary accretion. We developed sixty N-body simulations to describe the dynamical processes involved during and after the migration of the gas giant.
}
{Our simulations produce three different classes of planets in the HZ: `water worlds', with masses between 2.75 $M_{\oplus}$ and 3.57 $M_{\oplus}$ and water contents of 58 \% and 75 \% by mass, terrestrial-like planets, with masses ranging from 0.58 $M_{\oplus}$ to 3.8 $M_{\oplus}$ and water contents less than 1.2 \% by mass, and `dry worlds', simulations of which show no water.
A relevant result suggests the efficient coexistence in the HZ of a Jupiter-mass planet and a terrestrial-like planet with a percentage
of water by mass comparable to the Earth. Moreover, our study indicates that these planetary systems are dynamically stable
for at least 1 Gyr.
}
{Systems with a Jupiter-mass planet located at 1.5 au - 2 au around solar-type stars are of astrobiological interest.
These systems are likely to harbour terrestrial-like planets in the HZ with a wide diversity of water contents.
}


\authorrunning{ L. A. Darriba et al.}
\titlerunning{Earth-like planets with migrating Jupiter up to the habitable zone}

\maketitle
%----------------------------------------------------------------------
\section{Introduction}

In recent years, observational works \citep{Cumming2008,Howard2013} and theoretical studies \citep{Mordasini2009,Ida2013}
have suggested the existence of a wide diversity of planetary
architectures in the Universe. The number of confirmed planets discovered to date outside the solar system has grown to 3610, of which 1192 are associated with G-type stars. From this group, $\sim$25\% are Jupiter-mass planets, covering a wide range of semi-major axes, which extend from 0.01 au to 6 au.

Gaseous giants are thought to be formed in the outer regions of a protoplanetary disk (e.g. \citet{Pollack1996}). However, the observational evidence shows that a great number of giant exoplanets discovered around Sun-like stars are located at inner regions of the system, well inside the position of the so-called snowline at $\sim$ 3 au. A peculiar kind of such `warm-giant planets' is represented by those orbiting within 0.1 au from the central star, which are known as `hot Jupiters'. A possible explanation for the origin of such warm and hot Jupiters
suggests that they formed by gradual accretion of solids and capture of gas at a larger distance from the star, and it subsequently migrated inwards through interactions with the gas disk, from a process known as `type II migration' \citep{Lin1996}.

Several authors have analysed the formation, evolution, and survival of terrestrial-like planets during and after
gaseous giant migration. On the one hand, \citet{Armitage2003} suggested a reduced efficiency of terrestrial-like planet formation
following gas giant migration. In such a work, the author made use of a simple time dependent model for the evolution
of gas and solids in the disk and inferred that replenishment of solid material in the inner disk, following the inward
migration of a giant planet, is generally an inefficient process. From this, \citet{Armitage2003} suggested that terrestrial planets
are conceivable to survive in a modest fraction of systems in which a single generation of massive planets formed without significant
migration. However, it is worth noting that the author did not consider the dynamical effects of the migrating giant
on any planetesimals. On the other hand, \citet{Mandell2003} developed numerical simulations in order to analyse
how the inward migration of a gas giant affects the evolution of inner terrestrial planets. To do this, the authors placed
a planet identical to Jupiter at 5.2 au, which was allowed to migrate inward over different timescales through a inner
terrestrial-like planet system. From this, they concluded that the inward migration of a giant planet does not invariably
eliminate pre-formed terrestrial planets and that, given an initial layout of bodies similar to that of the solar system,
between 1\%-4\% of systems in which migration occurred could still possess a planet in the habitable zone (HZ). It is very important
to take into account that none of these studies actually simulated the terrestrial planet formation process simultaneously
with the giant planet migration.

\citet{Fogg2005} developed the first work aimed at studying the inner system planetary accretion in the presence of a
migrating giant. From N-body simulations, this study examined the oligarchic and giant-impact growth in a protoplanet or planetesimal
disk extending between 0.4 au and 4 au. Five scenarios were considered, corresponding to five different ages for the inner planet
forming disk at the point when a giant planet was assumed to form at 5 au and migrate in to 0.1 au. In all five of their scenarios,
\citet{Fogg2005} found that the majority of the disk solids survived the passage of the giant planet, either by being shepherded
inward of the giant, or by being scattered by the giant into excited exterior orbits. Then, \citet{Fogg2007} studied
the formation of terrestrial-like planets in hot-Jupiter systems, including an improved model that incorporated a viscous evolving
gas disk, annular gap and inner-cavity formation due to the gravitational field of the giant planet, and self-consistent evolution of
the giant's orbit. From this, the authors obtained that more than 60\% of the embryos and planetesimals survives the migration
of the giant planet by being scattered by it into external orbits, which results in the regeneration of a solid disk where it is possible for the
formation process of terrestrial-like planets to occur. \citet{Fogg2007} inferred that the giant planet migration induces
mixing of volatile-rich material from beyond the called snowline into the inner region of the disk, suggesting that hot-Jupiter systems
are likely to harbour water-abundant terrestrial-like planets in their circumstellar HZs. It is worth remarking that
in the works developed by \citet{Fogg2005, Fogg2007} the authors analysed the evolution of the disk and the formation process of terrestrial-like planets
for a few Myr after the migration of the gaseous giant.

\citet{Mandell2007} carried out N-body simulations with the goal of exploring the formation of terrestrial-like planets during and after
giant planet migration, analysing the dynamics of post-migration planetary systems over 200 Myr. These authors proposed models with a
single migrating giant planet, as well as with one inner migrating and one outer non-migrating giant planet. This work suggests that the
formation of terrestrial-like planets in the HZ seems to be a frequent process in such systems. However, an interesting result derived by
\citet{Mandell2007} indicates that planets surviving in the HZ in systems with close-in giant planets tend to accrete more water-rich
material than those in systems with only outer gas giants.

Then, \citet{Fogg2009} analysed the formation and evolution of terrestrial-like planets in warm-Jupiter systems. In particular, the
authors studied the evolution of planetary systems in which a gaseous giant halts migration at semi-major axes in the range 0.13 au - 1.7 au
due to the dissipation of the gas disk. A very interesting result derived by these authors suggests that terrestrial-like planets can
survive in the HZ if the giant planet's final orbit lies outside the region of $\sim$ 0.3 au-1.5 au for a 1 $M_\odot$ star.

Here, we present results of numerical simulations aimed at analysing the effects generated by a migrating gaseous giant
around a Sun-like star. In particular, our study focusses on the formation of terrestrial-like planets and water delivery in the HZ
in systems in which a Jupiter-mass planet halts migration at semi-major axes in the range 1.3 au-2 au due to gas disk dispersal.
Among the works previously mentioned, that of \citet{Fogg2009} is the most similar to that shown in this paper, though
there are several relevant differences that deserve to be discussed. In fact, the main differences between the present paper and that
carried out by \citet{Fogg2009} are related to the mass of the protoplanetary disk, the size of the planetesimals, the physical
and orbital properties of the giant planet, and the initial distributions of planetary embryos and planetesimals to be used in the
N-body simulations. Moreover, unlike \citet{Fogg2009}, we propose to analyse the long-term dynamical stability of the resulting
systems, integrating the simulations of interest for a time span of 1 Gyr. In particular, we believe our model improves the
initial conditions assigned to embryos and planetesimals for the development of the N-body simulations in comparison with those proposed
by \citet{Fogg2009}. This point of our research is very important because a more realistic treatment of the initial conditions
associated to embryos and planetesimals will lead to a better determination of the physical and dynamical properties of the resulting
planets. Moreover, our research offers a more detailed treatment concerning the long-term dynamical stability.
This investigation will allow us to strengthen our knowledge concerning the astrobiological interest of planetary systems around Sun-like
stars that host a Jupiter-mass planet at intermediate distances.

This paper is therefore structured as follows. The properties of the protoplanetary disks used in our study are presented in Sect. 2.
In Sect. 3, we discuss the semi-analytic model that allows us to describe the formation of the giant planet and the evolution of
the system in the gas phase. Moreover, this model allows us to outline our choice of initial conditions for the N-body simulations.
Then, we describe the main features of the N-body code used by us in Sect. 4. In Sect. 5, we show results and carry out
a detailed analysis of all simulations. Finally, we carry out a discussion of these results and expose our conclusions in Sect. 6.

\section{Properties of the protoplanetary disk}
\label{section_protoplanetary_disk}

In order to determine the distribution of material in a protoplanetary disk, the most relevant parameter is the surface density.
The surface density profiles of gas ($\Sigma_{\text{g}}$) and planetesimals ($\Sigma_{\text{p}}$) adopted in our model are the same
as those corresponding to the classical Minimum Mass Solar Nebula (MMSN) derived by \citet{Hayashi1981}:
\begin{eqnarray}
\Sigma_{\text{g}}(R) &=& 1700 \left(\frac{R}{1~\text{au}} \right)^{-3/2}~\text{g}~\text{cm}^{-2}, \\
\Sigma_{\text{p}}(R) &=& \left \lbrace \begin{array}{ll}
7.1 \left(\displaystyle{\frac{R}{1~\text{au}}}\right)^{-3/2}~\text{g}~\text{cm}^{-2}, & \mbox{if $R < 2.7$~au},\\
\\
30 \left(\displaystyle{\frac{R}{1~\text{au}}}\right)^{-3/2}~\text{\
g}~\text{cm}^{-2}, & \mbox{if $R > 2.7$~au},
\end{array}
\right.
\label{e1}
\end{eqnarray}
\noindent{where} $R$ represents the radial coordinate in the mid-plane, and the discontinuity at 2.7 au in the planetesimal surface density is caused by the condensation of volatiles, particularly the condensation of water (often called snow line). 

In fact, for our work we have adopted a disk N times more massive than the MMSN in order to trigger the formation of a giant planet just beyond the snow line. The numerical value of the scalar N is accurately chosen using a semi-analytic model, which will be explained in detail in Sect. \ref{section_semianalitical_model}.

In our model we have assumed a central solar-type star of 1 $M_{\odot}$ and solar metallicity. Other important parameters, such as the mass
of the protoplanetary disk $M_{\text{d}}$, and the gas dissipation timescale $\tau$ must be quantified in order to narrow
down the wide spectrum of possible scenarios for our simulations. \citet{Mamajek2009} showed that the distribution
of disk lifetimes decays exponentially with a
timescale of $\sim 2.5$ Myr, and \citet{Alexander2006} and \citet{Armitage2010} showed that after a few Myr of viscous evolution, the disk is completely dissipated in a timescale of $\sim 10^5$ yr, when photoevaporation is included. Since the photoevaporation timescale is a factor $\sim 20$ smaller than the disk lifetime, in our work we have considered that the photoevaporation is instantaneous.

In order to define the values of $M_{\text{d}}$ and $\tau$, it is
necessary to develop a detailed study about the evolution of the protoplanetary disk during the gas phase with the aim of forming a giant
planet beyond the snow line. In this case, we want to analyse the dynamical evolution of a planetary system with a gas giant migrating
inward. In particular, we considered that the formation of the gaseous giant is a late process, so that its migration should be
truncated at intermediate distances due to the dissipation of the gaseous disk.

To specify which protoplanetary disks can lead to this scenario, we used a semi-analytic model with which we were able to analyse the
evolution of a planetary system in the gaseous phase. As we will see in the next section, this model allowed us to define the
parameters $M_{\text{d}}$, and $\tau$ as well as the initial conditions concerning the distribution of embryos and planetesimals
to be used later in the N-body simulations.

%----------------------------------------------------------------------
\section{Semi-analytic model for the protoplanetary disk}
\label{section_semianalitical_model}

\subsection{Description}
\label{section_semianalitical_model_description}

Our model to calculate the formation of the planetary system and the giant planet is the one described in \citet{Guilera2010}, \citet{Guilera2011} and \citet{Guilera2014}. Here, we discuss the most important physical properties
of the model, which calculates the formation of a planetary system immersed in a protoplanetary disk that evolves in time.

The protoplanetary disk is characterised by a gaseous and a solid component (a planetesimal disk). On the one hand, the gaseous component is dissipated
exponentially,
\begin{eqnarray}
\Sigma_{\text{g}}(t)= \Sigma_{\text{g}}(t=0)\exp(-t/\tau),
\label{eq:evol_Sg}
\end{eqnarray}
where $\tau$ is a characteristic timescale. On the other hand, the planetesimal disk evolves by planetesimal migration due to nebular
drag and planetesimal accretion by the embryos (in this work we do not include the collisional evolution of the population of
planetesimals). As a consequence of the mass conservation, the planetesimal surface density obeys a continuity equation,
\begin{eqnarray}
\frac{\partial \Sigma_{\text{p}}(R,r_{\text{p}})}{\partial t} - \frac{1}{R} \frac{\partial}{\partial R} \Big[ Rv_{\text{mig}}(R,r_{\text{p}})
\Sigma_{\text{p}}(R,r_{\text{p}}) \Big] = \mathcal{F}(R,r_{\text{p}}),
\label{eq:evol_Sp}
\end{eqnarray}
where $r_{\text{p}}$ represents the planetesimal radius, $v_{\text{mig}}$ the planetesimal migration velocity and $\mathcal{F}$ represents the
sink terms due to the accretion by the embryos.

Two major processes govern the evolution of the eccentricities and inclinations of the planetesimal population along the disk:
\begin{itemize}
\item[$\bullet$] the gravitational stirring produced by the embryos \citep{Ohtsuki2002},
\item[$\bullet$] the damping due to nebular gas drag \citep{Rafikov2004,Chambers2008}.
\end{itemize}
The gas drag also causes an inward orbital migration of the planetesimals. The interaction between the planetesimals and the gas
depends on the planetesimal relative velocities with respect to the gas, and on the ratio between the planetesimal radius and the gas
molecular mean free path. As in \citet{Guilera2014}, we considered three different regimes \citep{Rafikov2004,Chambers2008}:
\begin{itemize}
\item[$\bullet$] the Epstein regime,
\item[$\bullet$] the Stokes regime,
\item[$\bullet$] the quadratic regime.
\end{itemize}
Moreover, the embryos immersed in the disk grow by accretion of planetesimals in the oligarchic regime, and by the accretion of
the surrounding gas. For the accretion of planetesimals, we used the prescriptions given by \citet{Inaba2001}. Regarding the
accretion of the surrounding gas, we solved the classical equation of transport and structure for the planet envelope (see \citet{Fortier2009} and \citet{Guilera2010} for detailed explanations). Finally, when the distance between two
embryos becomes smaller than
3.5 mutual Hill radii, we considered the perfect merger between them. In this case, when two planets are merged we consider that
the mass of the new planet is the sum of the core mass of the planets that are merged (we considered that the envelopes of the
planets are dissipated), and the semi-major axis is the mean of the previous two semi-major axes (pondered by the mass of each planet).
It is important to remark that in this work we have not considered type I migration for the planets. We considered the evolution of the
planetary system taken into account only the in situ formation of the planets until one planet achieves the critical mass\footnote{As
critical mass we considered the mass of the planet core when the mass of the envelope equals it and the planet starts the gaseous
runaway growth.} to become a giant planet. Regarding type I migration, the studies of it in idealised isothermal disks predict rapid
inward migration rates \citep{Tanaka2002}, so it is necessary to reduce the migration rates using an ad-hoc factor to reproduce observations \citep{Alibert2005, Ida2008, Miguel2011a, Miguel2011b}. However, if more realistic disks are considered \citep{Kley2008, Paardekooper2010, Paardekooper2011} type I migration could substantially change.
Moreover, when the mass of the planet becomes of the order of $10 M_{\oplus}$,
corrotation torques become significant and type
I migration is no longer linear. In this regime, migration is slowed or reversed \citep{Masset2006, Dittkrist2014}.
But even in isothermal
disks type I migration could be outward if full MHD turbulence is considered \citep{Guilet2013}. Finally, in a recent work,
\citep{Benitez-Llambay2015} found that if the released energy by the planet due to
accretion of solid material is taken into account, this phenomena generates a heating torque which could significantly slow down,
cancel, and even reverse inward type I migration for planets with masses $\lesssim 5~\textrm{M}_{\oplus}$. 
In light of these recent results, we consider our assumption of in situ formation could be considered as an approximation of a
more complex problem.

\subsection{Application}
\label{section_semianalitical_model_application}

\begin{figure}[htb!]
\centering
\includegraphics[angle=-90, width= 0.48\textwidth]{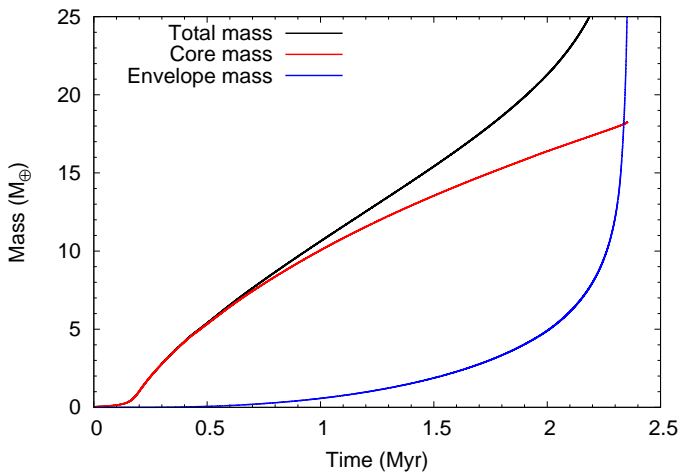}
\caption{Time evolution of the core mass, envelope mass, and total mass of a planet located just beyond the snow line ($\sim 3$~au).
Chosen a disk $\sim 1.35$ times more massive than the MMSN, and a characteristic timescale of $\tau= 2.5$ Myr for the dissipation of the
gas disk, a Jupiter-like planet is formed by the accretion of planetesimals with 100 m radius in 2.35 Myr.}
\label{fig:giant_evolution}
\end{figure}

As we mentioned before, we applied our model to calculate the formation of a planetary system until one planet achieves the critical
mass and becomes a giant planet.  We applied the same methodology as in \citet{deElia2013}. We determined the
mass of the disk to be used in our simulations calculating the in situ formation of an embryo located just beyond the snow line
($\sim 3$ au) using our model of giant planet formation \citep{Guilera2010}. In this work, we have considered a particular scenario wherein the formation of a giant planet is assumed to happen just before the dissipation of the disk. Chosen a characteristic timescale of $\tau= 2.5$ Myr for the dissipation of the disk, and assuming instant photoevaporation, we find that our model allows us to form a Jupiter-like planet in 2.35 Myr using a disk $\sim 1.35$ times more massive than the MMSN, and using planetesimals with 100 m radius. This result is depicted in Fig.~\ref{fig:giant_evolution}, where we show the time evolution of the masses from both the core and the envelope of a giant planet
formed near the snow line ($\sim 3$~au). We note that in this simulation we used grain opacities without a reduction respect to the interstellar values. However, \citet{Ormel2014} and \citet{Mordasini2014} found that  grain opacities could be much lower than in the interstellar medium (ISM). Reducing the opacity due to grains to a value of 2\% of the ISM values \citep{Hubickyj2005} we found that, for the same disk, the cross-over mass is reached at $\sim 1$~Myr where the mass of the core at this time is $\sim 14 M_{\oplus}$. Another important assumption in our model of giant planet formation is that we did not consider a limitation in the accretion of gas. In order to analyse this assumption we performed a series of test-simulations using the code P{\scriptsize{LANETA}}LP \citep{Ronco2017}, where the limitation of the gas accretion onto the planet by the capability of the disk to supply enough material is considered, finding similar results if the Shakura-Sunyaev $\alpha$-parameter \citep{Shakura1973} takes values between $\alpha= 10^{-2} - 10^{-3}$. However, for small values of $\alpha$ ($\sim 10^{-4}$) gas accretion becomes significantly different. We want to remark that these assumptions could lead to a possible different set of initial conditions for the N-body simualtions (see below).  

\begin{figure}[htb!]
\centering
\includegraphics[angle=0, width= 0.48\textwidth]{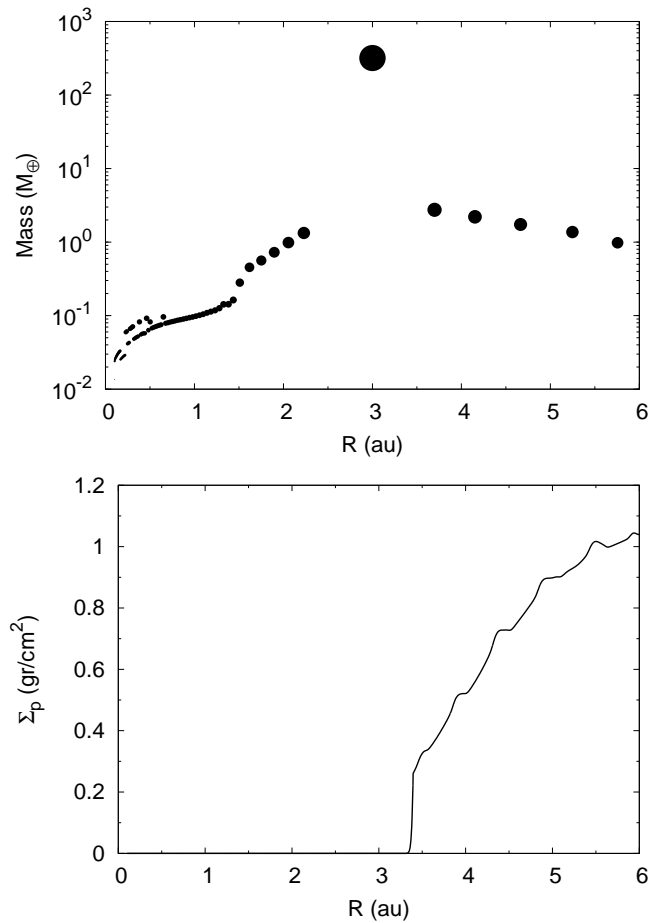}
\caption{Distribution of embryos (top) and planetesimals (bottom) when the Jupiter-like planet is formed. The point sizes represent
the embryos scaled in mass. We can see that the planetary embryos and the gas giant accreted all the planetesimals inside $\sim 3.5$~au.}
\label{fig:initial_conditions}
\end{figure}

Once the mass of protoplanetary disk was specified, we analysed the evolution of embryos and planetesimals during the gaseous phase.
In particular, our model evolves such populations inside and beyond the snow line, and the simulations stop when the giant planet is
formed. With this approach, we obtain the picture of the system when the giant planet is already formed and there is still a small
remnant of gas in the disk. To do this, we assumed that the initial mass of the embryos is the one corresponding to the beginning of
the oligarchic growth regime \citep{Ida1993}. Moreover, as we mentioned above, we considered planetesimals with 100 m radius
and a characteristic timescale of $\tau= 2.5$ Myr for the dissipation of the disk. Since the timescale for collisions between planetesimals of 100 m radius is smaller than the accretion timescale when the embryos are relatively small, we neglected the collisional evolution of planetesimals.
Figure~\ref{fig:initial_conditions} shows the distribution of embryos and planetesimals when the giant planet is formed.
In particular, the top panel represents the mass distribution of planetary embryos as a function of the distance from the central star,
while the bottom panel shows the surface density profile of planetesimals. These final distributions are used as initial condition
for the N-body simulations.

\section{N-body model}
\label{section_nbody}

\subsection{Characterisation and initial conditions}

The N-body code used to carry out our study is that developed by \citet{Chambers1999} and known as MERCURY. In particular,
we use the hybrid integrator, which uses a second-order mixed variable symplectic algorithm to treat the interaction between objects
with separations greater than 3 Hill radii, and a Bulirsch-St\"oer method for resolving closer encounters.

As we described in Sect. 3.2, we assumed that the gaseous component of the protoplanetary disk is dissipated in a timescale of 2.5 Myr,
while the formation of the giant planet is a late process occurring within 2.35 Myr. Thus, a small amount of gas remains on the disk
for 1.5 $\times$ 10$^{5}$ yr. In the present study, we analysed three distinct work scenarios, in which the gas giant migrates up
to different distances respect to the central star during the 1.5 $\times$ 10$^{5}$ yr after its formation. These scenarios are defined
as follows:
\begin{itemize}
 \item Scenario 1: the Jupiter-mass planet migrates up to $\sim 2$~au,
 \item Scenario 2: the Jupiter-mass planet migrates up to $\sim 1.6$~au,
 \item Scenario 3: the Jupiter-mass planet migrates up to $\sim 1.3$~au.
\end{itemize}

According to \citet{Tanaka1999}, a giant planet can act as a shepherd, when its migration timescale is greater than some critical timescale, defined as a function of the gas dissipation timescale, or as a predator, in the other case. By doing the calculations, we find that all the giants on the simulations are predators.\bigskip

In order to make use of the MERCURY code, it is mandatory to specify physical and orbital parameters for the gas giant planet,
the planetary embryos, and the planetesimals. In all three scenarios, we assumed one Jupiter-mass planet with a physical
density of 1.3 g cm$^{-3}$. Immediately after formation, the giant planet might have a radius a factor of two larger than its final state, thus increasing the collisional cross section. However, we considered that this increase would not alter the dynamics of the system significantly. This planet is assigned a circular and planar orbit (i.e. $e=0$ and $i=0$) with a semi-major axis of
3 au. Moreover, the longitude of the ascending node $\Omega$, the argument of the pericentre $\omega$, and the mean anomaly $M$
are randomly generated in the range $[0,360^{\circ}]$.

We consider $41$ planetary embryos between $0.5$ au and $6$ au in the three work scenarios. As we can see
in the top panel of Fig.~\ref{fig:initial_conditions}, the embryos have masses between $0.0135$ M$_{\oplus}$ and $2.75$ M$_{\oplus}$.
For all planetary embryos, we assumed a physical density of $3$ g cm$^{-3}$. Regarding the orbital parameters, eccentricities and
inclinations are randomly generated in the range $[0,0.02]$ and $[0,0.5^\circ]$, respectively. As for the giant planet,
the longitude of the ascending node $\Omega$, the argument of the pericentre $\omega$ and the mean anomaly $M$ are randomly generated
in the range $[0,360^{\circ}]$.

Finally, we consider 500 `super planetesimals', each one with a mass of $4.5 \times 10^{-3}$ $M_\oplus$ and a physical density of
1.5 g cm$^{-3}$. The semi-major axes of the super planetesimals are generated using the acceptance-rejection method developed by
John von Neumann, taking into account the surface density profile of planetesimals shown in the bottom panel of
Fig.~\ref{fig:initial_conditions}. Moreover, we assumed eccentricities and inclinations randomly generated in the range $[0,0.02]$
and $[0,0.5^\circ]$, respectively, and the values of $\Omega$, $\omega$ and $M$ are randomly generated in the range $[0,360^\circ]$
as in the case of embryos and the gas giant. The initial values for eccentricities and inclinations, both for embryos and planetesimals, were selected according to several authors who carried out N-body simulations, such as \citet{Raymond2004, Raymond2006, Raymond2009}, \citet{OBrien2006}, \citet{Walsh2011}, to cite a few.

The term super planetesimal is used to denote particles that represent a large number of small planetesimals. In these simulations, the Jupiter-mass planet and the planetary embryos feel the gravitational presence of all the particles in
the system, whereas the super planetesimals feel the effects of the central star and the big bodies, but they are non self-interacting.
This consideration is included to reduce the CPU time and to avoid that the high masses of the super planetesimals unrealistically
auto-excite the disk.

We assumed that the protoplanetary disk presents a radial composition gradient. In fact, given that the solid-surface density has
a jump past the snow line, we assumed that bodies beyond 2.7 au contain 75\% water by mass, while bodies inside 2.7 au do not have
water. This water distribution is assigned to planetary embryos and planetesimals of our simulations based on their starting locations.
An aspect to take into account is that the MERCURY code evolves the orbits of planetary embryos and planetesimals, allowing collisions to occur.
In all cases, collisions are treated as inelastic mergers preserving the mass and the water content.

Our main area of interest is the formation of planets inside the HZ. This zone represents a
region around the central star where a planet can retain liquid water on its surface.
\citet{Kopparapu2013a, Kopparapu2013b}
established inner and outer limits for the HZ around stars of different types. In particular, the authors define conservative
inner and outer edges of the HZ, which are determined by loss of water and by the maximum greenhouse provided by a CO$_{2}$ atmosphere,
respectively. For a Sun-like star, \citet{Kopparapu2013a, Kopparapu2013b} compute a conservative estimate for the width of the HZ
of 0.99-1.67 au. However, the authors also determine more optimistic limits for the HZ. For the inner edge, an optimistic estimate is
based on the inference that Venus has not had liquid water on its surface for at least the past one billion years \citep{Solomon1991}. For the outer edge, a more optimistic empirical limit can be estimated based on the observation
that early Mars was warm enough for liquid water to flow on its surface \citep{Pollack1987, Bibring2006}. For a Sun-like star, \citet{Kopparapu2013a, Kopparapu2013b} compute an optimistic estimate for the width of the HZ of 0.75-1.77 au. Given these estimates, we assumed that a planet is in the HZ of the system
and so can hold permanent liquid water on its surface if its whole orbit is contained inside the optimistic limits, that is,
if it has a perihelion $q \geq$ 0.75 au and an aphelion $Q \leq$ 1.77 au.

However, it seems too conservative to require that the perihelion and aphelion distances are both inside the HZ for a planet to be
habitable. \citet{Williams2002} showed that, provided that an ocean is present to act as a heat capacitor, it is
primarily the time-averaged flux that affects the habitability over an eccentric orbit. Planets with high orbital eccentricities
(e $>$ 0.1) have higher average orbital flux, which may help eccentric planets near the outer edge of the HZ to maintain
habitable conditions. We will take this criterion into account for cases in which the planet has a aphelion on the outer edge of the HZ.

As the reader can see in both panels of Fig.~\ref{fig:initial_conditions}, we considered that our study region extends from 0.5 au to 6 au.
Then, we set the step-size of the integration to six days, which is shorter than $1/20$th of the orbital period of the innermost body
in the simulation. We integrated each simulation for at least 200 Myr, which is a good choice as an upper limit for the formation
timescale of the terrestrial planets of our solar system \citep{Touboul2007, Dauphas2011, Jacobson2014}. Given the stochastic nature of the accretion process, we carried out twenty different numerical simulations for each
of the three work scenarios.

The original MERCURY code was modified in order to incorporate some effects acting on the different system components. Remember that a
small amount of gas remains in the system during the first 1.5 $\times$ 10$^{5}$ yr of evolution. From this, we considered the effects of
the type II migration on the Jupiter-mass planet. Then, we incorporated the effects of damping over the planetary embryos due to the
presence of the gas. Finally, we added the effects of drag and damping over the planetesimals due to the action of the gaseous component.
The inclusion of such effects is described in the following.

%--------------------------------------------------
\subsection{Giant planet migration}
\label{section_nbody_migration}

In order to model the type II migration, we considered a very simple drag acceleration, as described in \citet{Chiang2002}),
assuming that the giant planet migrates with a constant velocity, given by the expression
\begin{equation}
 \mathbf{a}_{\text{mig}} = -\frac{\mathbf{v}}{\tau_{\text{mig}}},
\end{equation}
\noindent{where} $\mathbf{v}$ is the velocity vector of the body, and $\tau_{\text{mig}} = $ 1.5 $\times$ 10$^{5}$ yr is the migration time
of the gaseous giant.

\subsection{Effects over planetary embryos: eccentricity and inclination damping}
\label{section_nbody_effects_over_embryos}

The eccentricities $e$ and inclinations $i$ of planetary embryos are damped due to interactions with the gaseous disk. According to \citet{Cresswell2008}, the eccentricity and inclination damping times $t^{\text{d}}_e$ and $t^{\text{d}}_i$, respectively,
are given by
\begin{equation}
\begin{array}{ccl}
t^{\text{d}}_e &=& \frac{t_{\text{wave}}}{\text{0.780}} \\ 
               &\times& \left[1-\text{0.14}\left(\frac{e}{H/R}\right)^{2}+\text{0.06}\left(\frac{e}{H/R}\right)^{3} +\text{0.18}\left(\frac{e}{H/R}\right)\left(\frac{i}{H/R}\right)^{2}\right], \\
t^{\text{d}}_i &=& \frac{t_{\text{wave}}}{\text{0.544}} \\
               &\times& \left[1-\text{0.30}\left(\frac{i}{H/R}\right)^{2}+\text{0.24}\left(\frac{i}{H/R}\right)^{3} +\text{0.14}\left(\frac{e}{H/R}\right)^{2}\left(\frac{i}{H/R}\right)\right], 
\label{eq:damping}
\end{array}
\end{equation}
\noindent{where} $t_{\text{wave}}$ is given by
\begin{equation}                                                                                                                                                            
 t_{\text{wave}} = \frac{M_\star}{m_{\text{e}}} \frac{M_\star}{\Sigma_{\text{g}} a^2_{\text{e}}} \left(\frac{H}{R}\right)^4 \Omega^{-1},                         
\end{equation} 
\noindent                                                                                                                                                                   
being $M_\star$ the stellar mass, $m_{\text{e}}$ and $a_{\text{e}}$ the mass and semi-major axis of the planetary embryo, respectively,
$\Sigma_{\text{g}}$ the gas surface density, $\Omega$ the angular velocity of the unperturbed disk, and $H/R$ is the disk's scale
height-to-radius ratio. The parameter $H$ is given by
\begin{equation}
 H = \left(\frac{K T R^3}{GM_\star m}\right)^{1/2},
 \label{eq:H}
\end{equation}
\noindent
being $m$ the mass of an hydrogen molecule, the radial temperature profile T is $T=280R^{-1/2}$ %$
(where R is given in au and T in Kelvin) and $K$ the Boltzmann constant.

Thus, the accelerations experienced by the planetary embryos due to the gaseous disk are written by
\begin{equation}
 \begin{array}{lll}
 \mathbf{a}^{\text{d}}_e &=& -2 \dfrac{\mathbf{v}\cdot\mathbf{r}}{r^2 t^{\text{d}}_e} \mathbf{r} \\
 \mathbf{a}^{\text{d}}_i &=& -\dfrac{v_z}{t^{\text{d}}_i} \mathbf{\hat{k}},
 \end{array}
\end{equation}
\noindent{where} $t^{\text{d}}_e$ and $t^{\text{d}}_i$ are given by Eqs.~\eqref{eq:damping}, $\mathbf{r}$ and $\mathbf{v}$ are the position and velocity vectors
of the embryo, respectively, and $\mathbf{\hat{k}}$ is the unity vector pointing in the direction perpendicular to the mid-plane.

%--------------------------------------------------
\subsection{Gas effects over planetesimals: inward migration and eccentricity and inclination damping}
\label{section_nbody_gas_drag}

Regarding the gas effect over the super-planetesimals, they experience a drag from the gas disk, where we consider a defined physically
realistic planetesimal size. For simplicity purposes, each planetesimal is considered as a sphere of radius $r_{\text{p}}$ and density $\rho_{\text{p}}$.
The expression for the gas drag is given by
\begin{equation}
  \mathbf{a}_{\text{drag}} = -\frac{3 C_{\text{D}} \rho_{\text{g}}}{8 r_{\text{p}} \rho_{\text{p}}} v_{\text{rel}} \mathbf{v}_{\text{rel}},
\end{equation}
\noindent
\citep{Adachi1976} where $r_{\text{p}}$ and $\rho_{\text{p}}$ are the radius and density of each individual planetesimal, respectively, $C_{\text{D}}$ is
a constant, which we define as $C_{\text{D}}=1$, $\mathbf{v}_{\text{rel}}$ is the vector velocity of the planetesimals relative to the gas, and $\rho_{\text{g}}$
is the volume gas density, given by
\begin{equation}
 \rho_{\text{g}} = \rho_0 e^{-{z^2/\mathbf{2}H^2}} e^{-t/\tau},
\end{equation}
\noindent
where $z$ is the coordinate normal to the mid-plane, and $\rho_0$ a constant given by
\begin{equation}
 \rho_0 = \frac{\Sigma_{\text{g}}}{H\sqrt{\mathbf{2}\pi}},
\end{equation}
\noindent
being $H$ is the height scale defined by Eq. \eqref{eq:H}, and $\tau$ the gas dissipation timescale specified in
Sect. \ref{section_protoplanetary_disk}. The value chosen for $r_{\text{p}}$ is 100~m (see Sect. \ref{section_semianalitical_model_application}). The planetesimals therefore experience both an orbital damping effect and an inward migration, decreasing with mass and increasing
with gas density.

\section{Results}

In this section we show the results of the simulations corresponding to the three scenarios mentioned before. In particular, we are
interested in analysing the formation of terrestrial planets in the HZ, focussing on their final masses, water contents, and long-term
dynamical stability.

\subsection{Scenario 1}
\label{sect:scenario1}

Here, we analyse the dynamical evolution of the systems in which the Jupiter-mass planet migrates up to $\sim$2 au from the central
star.
This set of runs represents a complement of the work developed by \citet{Fogg2009}, who analysed the terrestrial-like planet
formation in systems in which a giant planet halts migration at semi-major axes in the range 0.13 au - 1.7 au.
In this Scenario 1, six of the twenty N-body simulations form planets in the HZ of the system.
In Table \ref{table:resultados-Scen1} we have listed the simulations which end with a planet in the HZ after 200 Myr of evolution.
For each planet, we detailed the semi-major axis at the beginning and at the end of the integration, together with the final mass and the final percentage of water by mass. As the reader can see, the planets formed in the HZ are super-Earths with a very
wide range of final water contents.

\begin{table}[ht]
\caption{
 Planets surviving in the HZ after 200 Myr of evolution for Scenario 1.
 $a_{\text{i}}$ and $a_{\text{f}}$ represent the initial and final semi-major axes in au, respectively, $M$ the final mass
 in $M_\oplus$, and $W$ the final percentage of water by mass after 200 Myr of evolution.}
 \begin{center}
  \begin{tabular}{|c|r|r|r|r|}
 \hline
 Simulation & $a_{\text{i}}$ (au) & $a_{\text{f}}$ (au) & $M$ ($M_{\oplus}$) & $W$ (\%) \\
 \hline \hline
 3  & $2.23$ & $0.90$ & $2.15$ & $0.00$ \\
 7  & $2.06$ & $0.95$ & $3.85$ & $0.53$ \\
 11 & $2.23$ & $1.09$ & $2.40$ & $0.43$ \\
 14 & $3.70$ & $1.10$ & $2.75$ & $75.00$ \\
 16 & $2.23$ & $0.97$ & $3.62$ & $0.09$ \\
 17 & $3.70$ & $1.07$ & $3.57$ & $58.05$ \\
 \hline
  \end{tabular}
 \label{table:resultados-Scen1}
 \end{center}
\end{table}

\begin{figure*}[htb!]
 \centering
\includegraphics[angle=0, width= 0.98\textwidth]{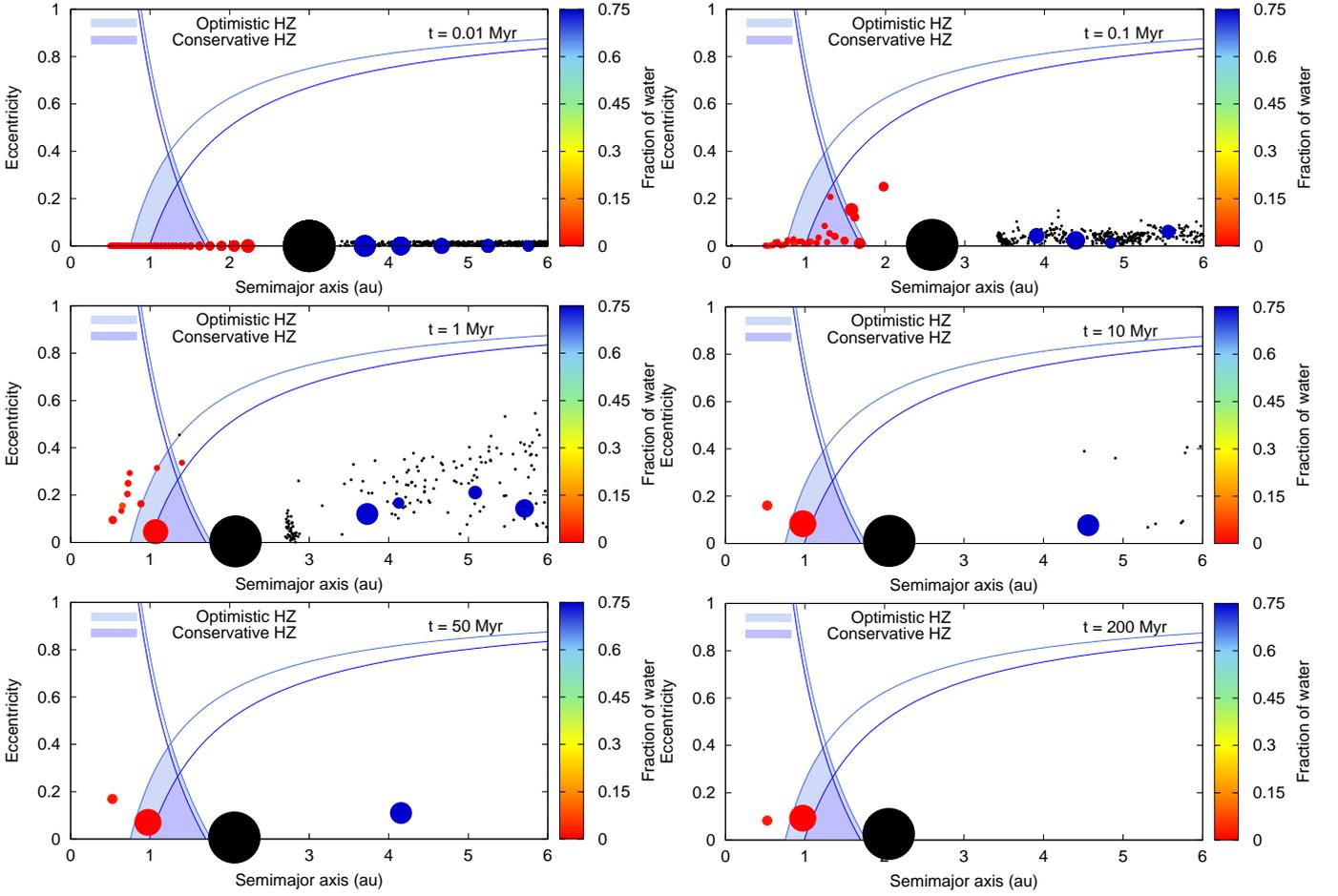}
 \caption{
 Snapshots of the dynamical evolution for Simulation 16 of Scenario 1. The blue and light-blue shaded areas represent the conservative
 and the optimistic HZ, respectively. In the same way, the blue and light-blue curves represent curves of constant perihelion and
 aphelion, both for the conservative and the optimistic HZ. The Jupiter-mass planet is plotted as a big black circle, the planetary
 embryos as coloured circles, and the planetesimals as small black points. The colour scale represents the fraction of water of the
 embryos respect to their total masses. Colour figure is only available in the electronic version.
 }
 \label{fig:time-evolution-Scen1-Sim16}
\end{figure*}

\begin{figure*}[htb!]
 \includegraphics[width=0.98\textwidth]{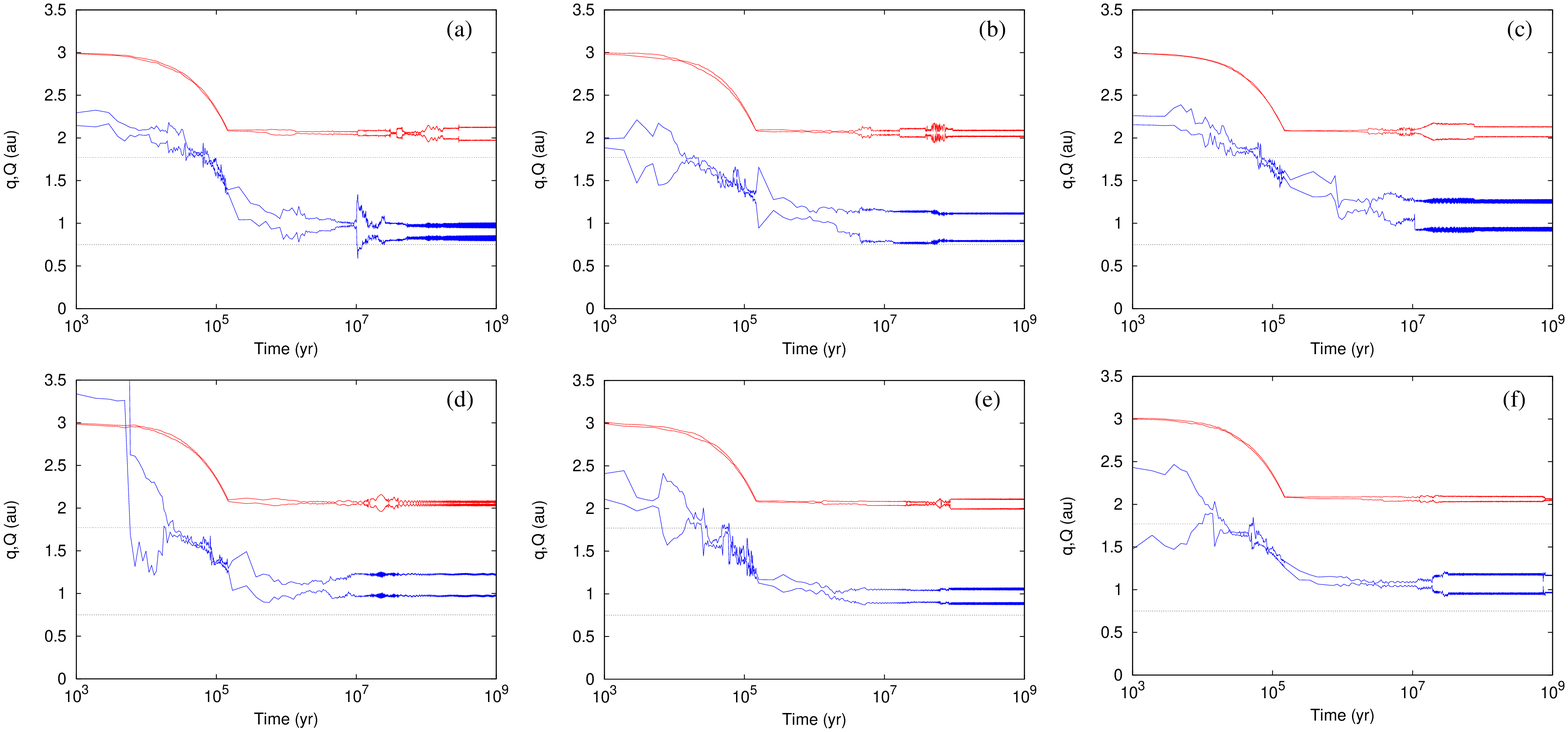}
 \caption{
 Evolution in time of perihelion and aphelion for the Jupiter-mass planet (red curve) and the planet surviving in the HZ (blue curve). These results, obtained from Simulations 3, 7, 11, 14, 16, and 17 of Scenario 1, are shown in the panels a), b), c), d), e), and f), respectively.
 The horizontal dashed lines represent the inner and outer edges of the optimistic HZ. These planetary configurations remain dynamically stable for 1 Gyr.}
\label{fig:qQ_vs_t_Scen1_Sim16}
\end{figure*}

The dynamical evolution of the six simulations that produce planets in the HZ is very similar. Thus, we analysed the results of
one of them as a representative of the whole group. The simulation proposed is that labelled as Simulation 16, and is shown in
Fig. \ref{fig:time-evolution-Scen1-Sim16}. In each panel of this figure we can see snapshots at different times of
the evolution on the semi-major axis-eccentricity plane.
In Fig. \ref{fig:time-evolution-Scen1-Sim16}, both embryos and planetesimals are quickly excited, reaching eccentricities of $e \sim 0.5$
in $1$ Myr, together with the ejection of $\sim 40\%$ of the planetesimals. At 10 Myr, $\sim 75\%$ of planetesimals are ejected, some
embryos grow accreting other embryos, and others collide with the giant, leaving three super-Earths, with masses ranging from 1 $M_\oplus$
to 3 $M_\oplus$.
One of these planets is located inside the HZ, and remains there until the end of the integration. In fact, the embryos distribution
barely change in the rest of the evolution.
At 50 Myr, only $\sim 5\%$ of planetesimals remain, and they are outside our interest zone. Finally, between 50 Myr and 200
Myr, the furthest embryo was pushed beyond $6$ au. Thus, at the end of the simulation, only two planets remain in our area of study together with the gas giant.
The innermost planet has a mass of $0.8$ $M_\oplus$ and the one inside the HZ has a mass of 3.62 $M_\oplus$, with a water content of 
$\sim 0.09\%$ by mass, which corresponds to $\sim 12$ Earth oceans\footnote{The amount of water on Earth's surface is 2.8
$\times$ 10$^{-4}$ $M_{\oplus}$, which represents 1 Earth oceans. On the other hand, \citet{Marty2012} suggested that the
current water content in Earth's mantle is $\sim$2 $\times$ 10$^{-3}$ $M_{\oplus}$. From these studies, the current Earth might have
a water content of about 0.1\%-0.2\% by mass.}. We also see that there is a near complete ejection of the initial planetesimals. In fact, $\sim$ 86$\%$
of them are ejected from the system, while only 1$\%$ remain outside our study region.

%In order to analyze the stability of the planets surviving in the HZ after 200 Myr of evolution, we extend those simulations shown in Table \ref{table:resultados-Scen1} up to 1 Gyr. The results indicate that the planets in the HZ do not suffer significant changes in their semimajor axes, eccentricities, and inclinations during 1 Gyr of evolution. We consider that this study suggests the long-term dynamical stability of the planets surviving in the HZ for Scenario 1.

If we look at the results shown in Table \ref{table:resultados-Scen1}, we can see that all planets surviving in the HZ end up the simulation with semi-major axes very similar to that corresponding to the Earth (i.e. $0.9$~au $\leq a \leq 1.1$~au). In order to analyse the potential habitability of such planets, we studied their orbital evolution for 1 Gyr.
This was done by computing the evolution in time of the orbit's innermost and outermost points (i.e. perihelion and aphelion)
for each planet surviving in the HZ.
Figure \ref{fig:qQ_vs_t_Scen1_Sim16} shows the evolution in time of the perihelion ($q$) and aphelion ($Q$) of the Jupiter-mass planet
and the planet surviving in the HZ for the six simulations of Scenario 1 shown in Table \ref{table:resultados-Scen1}
during 1 Gyr. For each simulation, the perihelion ($q$) and aphelion ($Q$) of the giant planet, shown in Fig. \ref{fig:qQ_vs_t_Scen1_Sim16}, decrease during its migration, remaining nearly constant after it stops. The values of $q$ and
$Q$ of the inner planet, which are also shown in Fig. \ref{fig:qQ_vs_t_Scen1_Sim16}, also decrease on this period,
falling inside the HZ, which is delimited by the dashed horizontal lines. This indicates that the planet surviving
in the HZ could retain liquid water on its surface, satisfying an essential requirement for the development and maintenance of life. We consider that this study suggests the long-term dynamical stability of the planets surviving in the HZ for Scenario 1.

\begin{figure}[htb!]
 \centering
\includegraphics[angle=0, width= 0.48\textwidth]{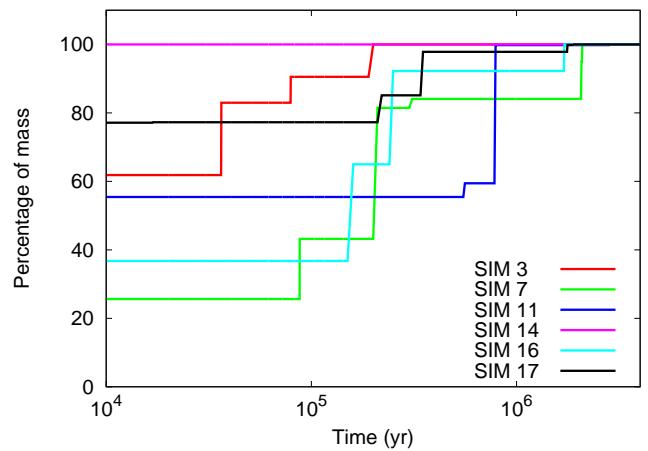}
 \caption{
 Mass of the planets surviving in the HZ obtained from Scenario 1 as a function of time.
 All these planets have formation timescales less than $\sim$ 2 Myr.
 }
 \label{fig:masa_t-Scen1}
\end{figure}

Another relevant information shown in Table \ref{table:resultados-Scen1} is the mass of each planet surviving in the HZ after
200 Myr of evolution. In all these simulations, the planets in the HZ are super-Earths with masses ranging from
2.15 $M_\oplus$ to 3.85 $M_\oplus$.
Figure \ref{fig:masa_t-Scen1} shows the percentage of mass of each planet of Scenario 1 surviving in the HZ as a function of time.
With the exception of the super-Earth formed in the Simulation 14, which does not collide with any other body during the entire
integration, the rest of the planets acquire their final masses in less than 2 Myr. According to that suggested by \citet{Jacobson2014}, this represents a timescale shorter than that associated to the formation of the Earth. In agreement with the
generally accepted scenario, the last giant impact on Earth formed the Moon and initiated the final phase of core formation by melting
Earth's mantle. Using highly siderophile element abundance measurements, \citet{Jacobson2014} determined a Moon-formation
age of 95$\pm$32 Myr after the condensation of the first solids in the solar system.

\begin{figure*}[htb!]
 \centering
\includegraphics[angle=0, width= 0.98\textwidth]{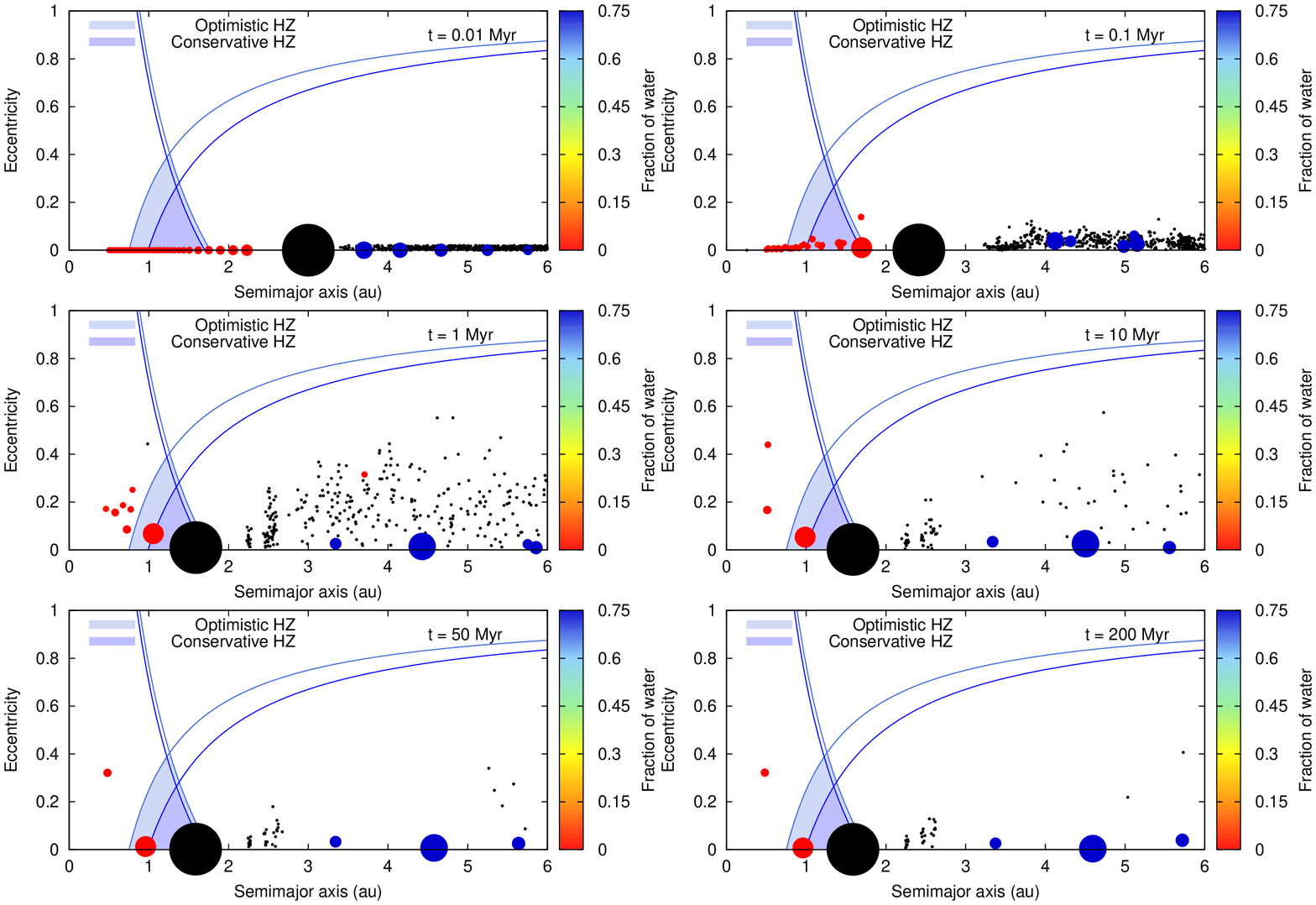}
 \caption{Snapshots of the dynamical evolution for Simulation 6 of Scenario 2. The colour code description is analogous to that of Fig. \ref{fig:time-evolution-Scen1-Sim16}.}
\label{fig:time-evolution-Scen2-Sim15}
\end{figure*}

The last aspect we analyse is shown on the fifth column of Table \ref{table:resultados-Scen1}, which represents the water content
of each planet in the HZ at the end of the simulation. 
Our simulations produce three different kinds of planets in the HZ. The first kind is represented by planets
whose accretion seed\footnote{Following \citet{Raymond2009}, we define a planet's accretion seed as the larger body
in each of its collisions.} starts the simulation beyond the snow line. Such planets have significant primordial water contents and
end up the simulation with very high water percentages by mass. Thus, they are called water worlds (see \citet{Kuchner2003} and
\citet{Leger2004}, for instance, for further reference).
The second kind of planets formed in the HZ is represented by those whose accretion seed starts the simulation
inside the snow line and their feeding zones are primarily associated to the inner region of the disk. Such planets end the
simulation with water percentages less than $\lesssim$ 0.53\% by mass. In general terms, the planets of this kind formed in our simulations
are more massive than the Earth. However, such planets show features in their dynamical evolution and final water percentages
that result to be comparable to the Earth. Thus, we decide to refer to this second kind of planets formed in the HZ of our
simulations as `Earth-like planets'. We infer that such planets have a very important astrobiological interest since they
could achieve similar conditions to the Earth for the development and maintenance of life. Finally, the third kind of planets
formed in the HZ is represented by those whose accretion seed starts the simulation inside the snow line and do not accrete water-rich
material along the evolution. These planets end up the simulation with no water, and thus they are called dry worlds.

In this scenario, we can see from Table \ref{table:resultados-Scen1} that the planets in the HZ associated to Simulations 14 and 17,
which start their evolution at $a \sim$ 3.7 au, are truly water worlds. In fact, the planet in the HZ produced in Simulation 14
has a final mass of 2.75 $M_{\oplus}$ and a final water content of 75\%, which equals to 7366 Earth oceans. In the same way,
the planet in the HZ formed in Simulation 17 has a final mass of 3.57 $M_{\oplus}$ and a final water content of 58\%, which equals
to 7395 Earth oceans. The main difference between these two planets is that, since the one corresponding
to Simulation 14 does not suffer any collision during its evolution, it preserves the initial water proportion. On the contrary, the
planet in the HZ produced in Simulation 17 does experiment collisions with other bodies. In fact, this planet collides with three
dry embryos, which contribute to the final mass, and with three water-rich planetesimals, which contribute to the water content.
This phenomenon means that the planet in the HZ formed in Simulation 17 ends its evolution with a water percentage of 58\%
by mass. The final water contents of the water worlds formed in Scenario 1 are primarily determined
by their primordial water contents. Thus, the water acquisition for such planets is an early process since it is associated
to their formation.

On the other hand, the planets in the HZ associated to Simulations 7, 11 and 16 start the evolution with a semi-major axis inner
to the snow line, so they do not contain initial water. These planets accrete water-rich planetesimals
during their evolution, which play a primary role in such aspect. In fact, a significant result suggests that the planetesimals
represent the main source of water of the Earth-like planets formed in Simulations 7, 11 and 16. Such as the
Table \ref{table:resultados-Scen1} shows, the planet in the HZ formed in Simulation 7 has a final mass of 3.85 $M_{\oplus}$ and
a final water content of 0.53\%, which equals to $\sim$73 Earth oceans. In the same way, the planet in the HZ formed in the
Simulation 11 has a final mass of 2.4 $M_{\oplus}$ and a final water content of 0.43\%, which equals to $\sim$37 Earth oceans.
Finally, the planet in the HZ associated to Simulation 16 has a final mass of 3.62 $M_{\oplus}$ and
a final water content of 0.09\%, which equals to $\sim$12 Earth oceans. These three planets reach
75\% of their final water contents in times of the order of their formation time. Thus, the water delivery on them is
a late process in their evolutionary histories.

Finally, the planet in the HZ produced in Simulation 3 results to be a dry world. In fact, this planet ends the integration with
no water, since it collides with planetary embryos formed inner to the snow line.

A relevant result derived from the present research indicates that a Jupiter-mass planet orbiting at $\sim$2 au around a Sun-like star
is able to coexist with super-Earths associated to the HZ with a wide diversity of water contents. It should be noted that, although none of the planets in the HZ is associated to mean motion resonances with the Jupiter-mass planet
at the end of the simulations, there exist temporal captures of several embryos during its migration. For instance, in Simulation 7, an embryo with an initial
semi-major axis of $\sim 1.43$ au is captured in a region near the 3:1 mean motion resonance with the Jupiter-mass planet up to
$10^{3}$ yr. Another example is, in the same simulation, an embryo with an initial semi-major axis of $\sim 2.23$ au, which is
captured in the 3:2 mean motion resonance with the giant planet up to $\sim 5 \times 10^4$ yr. The captures into several mean motion
resonances with the gas giant is a natural process in this kind of planetary systems.

\subsection{Scenario 2}
\label{sect:scenario2}

We studied the dynamical evolution of the systems in which the Jupiter-mass planet migrates up to $\sim$1.6 au from the central
star. In this particular scenario, four of the twenty N-body simulations produce planets in the HZ of the system after 200 Myr
of evolution. Table \ref{table:resultados-Scen2} shows the main properties of the planets surviving in the HZ at 200 Myr.
For each of such planets, we have specified the semi-major axis at the beginning and at the end of the integration, as well as
the final mass and the final percentage of water by mass.

\begin{table}[ht]
\caption{
 Planets surviving in the HZ after 200 Myr of evolution for Scenario 2.
 $a_{\text{i}}$ and $a_{\text{f}}$ represent the initial and final semi-major axes in au, respectively, $M$ the final mass
 in $M_\oplus$, and $W$ the  final percentage of water by mass after 200 Myr of evolution.
 }
 \begin{center}
  \begin{tabular}{|c|r|r|r|r|}
 \hline
 Simulation & $a_{\text{i}}$ (au) & $a_{\text{f}}$ (au) & $M$ ($M_{\oplus}$) & $W$ (\%) \\
 \hline \hline
 1  & $0.50$ & $0.94$ & $0.58$ & $1.18$ \\
 6  & $2.06$ & $0.96$ & $3.51$ & $0.29$ \\
 14 & $2.23$ & $0.82$ & $3.50$ & $0.10$   \\
 15 & $2.23$ & $0.96$ & $1.71$ & $0.00$   \\
 \hline
  \end{tabular}
 \label{table:resultados-Scen2}
 \end{center}
\end{table}

Since all the simulations shown in Table \ref{table:resultados-Scen2} have an evolution which is dynamically comparable,
we consider the one labelled '6' as a representative of the group to describe their evolution.
Figure \ref{fig:time-evolution-Scen2-Sim15} shows snapshots in time on the semi-major axis-eccentricity plane of Simulation 6,
as we carried out in the previous section. The values of the time of each snapshot are the same as in
Fig. \ref{fig:time-evolution-Scen1-Sim16}.
We can see from Fig. \ref{fig:time-evolution-Scen2-Sim15} that at 1 Myr, the embryos inner to the giant planet
reach eccentricities $e \sim$ 0.25, while the planetesimals are excited acquiring $e \leq$ 0.6.
This excitation is possibly due to the proximity of the gaseous giant, since it ends its migration inside the HZ.
By 10 Myr more than 50\% of the planetesimals are ejected
from the system. In addition, six planetary embryos remain in our interest zone. 
During the rest of the evolution, most planetesimals are ejected. At 50 Myr the system reaches a planetary configuration that remains
without significant changes until the end of the evolution. At 200 Myr, the Jupiter-mass planet
and a super-Earth of 3.51 $M_{\oplus}$ with a water content of 0.29\% by mass coexist in the HZ. We note that $\sim$ 75 $\%$ 
of the initial planetesimals are ejected from the system, $\sim$ 6 $\%$ remain further than our interest zone, and $\sim$ 6 $\%$ with 
0.5 au $\leq a \leq$ 6~au.

\begin{figure*}[htb!]
 \includegraphics[width=\linewidth]{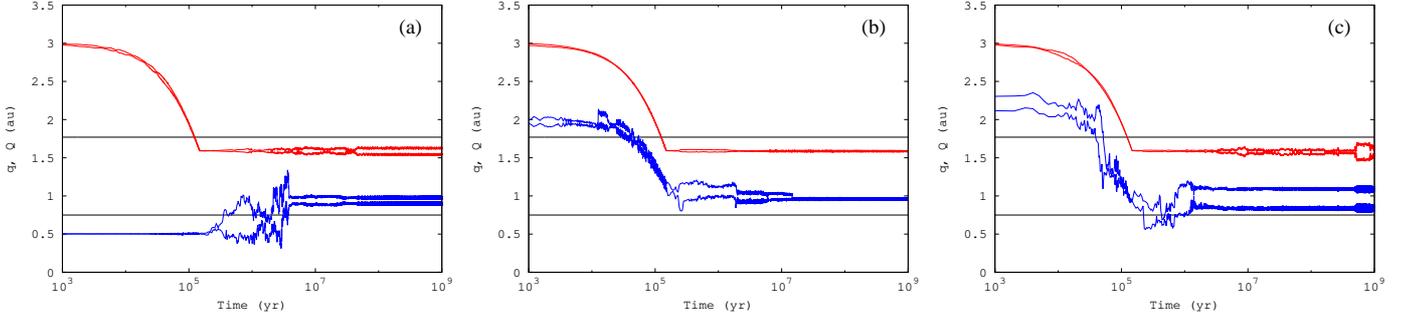}
 \caption{
 Evolution in time of perihelion and aphelion for the Jupiter-mass planet (red curve) and the planet surviving in the HZ (blue curve).
 These results, obtained from Simulations 1, 6 and 15 of Scenario 2, are shown in the panels a), b) and c), respectively. The horizontal
 dashed lines represent the inner and outer edges of the optimistic HZ. These planetary configurations remain dynamically stable for 1 Gyr.}
\label{fig:qQ_vs_t_Scen2}
\end{figure*}

\begin{figure*}[htb!]
 \includegraphics[width=0.98\linewidth]{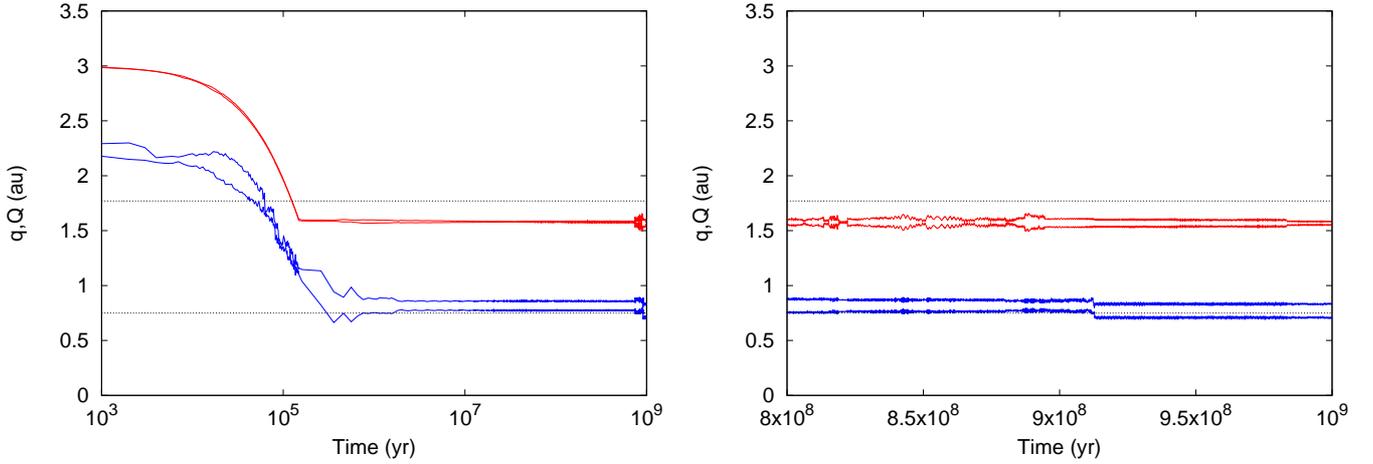}
  \caption{
 Evolution in time of perihelion and aphelion for the Jupiter-mass planet (red curve) and the planet surviving in the HZ (blue curve).
 The results displayed on the left panel correspond to Simulation 14 of Scenario 2. The right panel shows a zoom at the last 200 Myr
 of evolution, where a significant encounter shift the perihelion of the planet out of the inner edge of the HZ.}
\label{fig:qQ_vs_t_Scen2_Sim14}
\end{figure*}

In order to study the long-term dynamical stability of the planets surviving in the HZ at 200 Myr, we extend the four simulations
listed in
Table \ref{table:resultados-Scen2} for 1 Gyr. The planets associated to Simulations 1, 6, and 15 do not experience significant
changes in their semi-major axes, eccentricities, an inclinations during 1 Gyr of evolution. These three simulations end up with
a Jupiter-mass planet and a terrestrial-like planet inside the HZ. To ensure the potential habitability of such planets, we computed
the evolution in time of their perihelia and aphelia, which are shown in Fig. \ref{fig:qQ_vs_t_Scen2}. Our results indicate that
the planets in the HZ associated to Simulations 1, 6, and 15 keep their perihelia and aphelia inside the HZ for at least 1 Gyr.
This study helps us to conclude that such planets show a long-term dynamical stability and besides maintain habitability conditions
for at least 1 Gyr of evolution.
However, the planet in the HZ formed in Simulation 14 shows differences in the long-term dynamical behaviour. In fact, such a planet
has an encounter with another embryo at $\sim$920 Myr that produces relevant changes in its semi-major axis.
This dynamical event moves the perihelion of the planet outside the HZ, placing at to shorter distances from the central star than the
inner edge of the optimistic HZ. 
Figure \ref{fig:qQ_vs_t_Scen2_Sim14} shows the evolution of the perihelion and aphelion of the planet and the gas giant as a function
of the time. We can clearly see that the terrestrial-like planet experiments a shift in its orbit nearly at the end of the integration.
This phenomenon turns out to be crucial for the habitability, since a planet with such orbital properties
can not retain liquid water on its surface. Thus, the late dynamical instability produced in Simulation 14 significantly modifies
the orbit of the planet associated to the HZ, discarding it as potential target of astrobiological interest.
Consequently, using dynamical criteria, the group of runs with potentially habitable planets is reduced to Simulations 1, 6 and 15.

\begin{figure}[htb!]
 \centering
\includegraphics[angle=0, width= 0.48\textwidth]{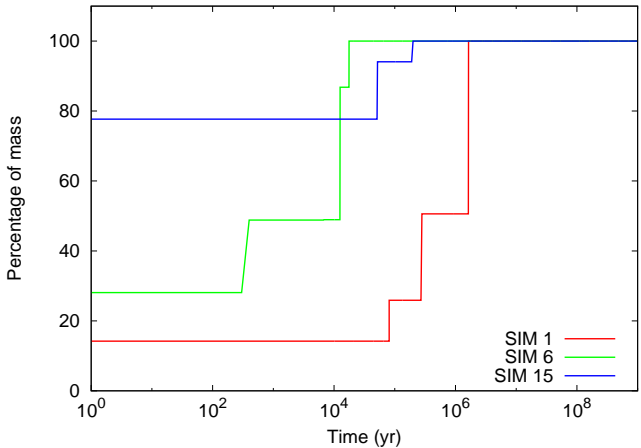}
 \caption{
 Mass of the planets surviving in the HZ obtained from Scenario 2 as a function of time.
 }
 \label{fig:masa_t-Scen2}
\end{figure}

As Table \ref{table:resultados-Scen2} shows, the planets associated to the HZ in Simulations 1, 6, and 15 have masses of
0.58 $M_\oplus$, 3.51 $M_\oplus$, and 1.71 $M_\oplus$, respectively. Moreover, Figure \ref{fig:masa_t-Scen2} shows the
percentage of mass of each of these planet surviving in the HZ as a function of time. This figure allows us to observe
that the planets in the HZ resulting from Simulations 1, 6, and 15 acquiring 90\% of their final mass at 1.6 Myr,
1.7 $\times$ $10^{4}$ yr, and 5 $\times$ $10^{4}$ yr, respectively, which represent timescales shorter than that associated
to the formation of the Earth \citep{Jacobson2014}.
The planet in the HZ from Simulation 1 has the lowest mass of the three, with a final value of 0.58 $M_\oplus$. This difference from the other two simulations is also reflected in their initial masses, with the Simulation 1 mass being about one order of magnitude smaller. Its initial
semi-major axis is also much smaller than the other two, being located near the internal border of the disk. All these features might
contribute to the explanation that the planet of the HZ associated to Simulation 1 takes $\sim 10^6$ years to form, a time two orders of magnitude longer than
the other two planets. 

Regarding the water content, none of the planets in the HZ formed in Simulations 1, 6, and 15
is a water world. In fact, as Table \ref{table:resultados-Scen2} shows, these three planets start the simulation at
distances from the central star closer than the snow line, so that they do not have primordial water contents. Moreover,
Table \ref{table:resultados-Scen2} indicates that such planets end the simulation with percentages of water less than $\sim$
1.2\% by mass.
In general terms, the three planets surviving in the HZ can be grouped into two different classes. On the one hand, the planet
resulting from Simulation 15 is a dry world. In fact, this planet does not undergo collisions with water-rich material during
its evolution, and thus it ends with no water at all. On the other hand, the planets in the HZ derived from Simulations 1 and 6
are Earth-like planets. In fact, as Table \ref{table:resultados-Scen2} shows, the planet resulting from Simulation 1
has a final mass of 0.58 $M_\oplus$ and a final water content of 1.2\% by mass, which is equal to $\sim$25 Earth oceans.
Our study indicates that this planet accretes its water content from the collision with an embryo at $\sim 1.6$ Myr,
which collided earlier with two outer water-rich planetesimals at $\sim 4 \times 10^4$ yr. In the same way,
the planet in the HZ formed in Simulation 6 has a final mass of 3.51 $M_\oplus$ and a final water content of 0.29\% by mass,
which is equal to $\sim$36 Earth oceans. This planets acquires its water content by colliding with an embryo at $1.7 \times 10^4$ yr,
which had earlier impacted with outer water-rich planetesimals. These planetesimals represent the main source
of water for the planets in the HZ formed in Simulations 1 and 6. Moreover, by looking at their water accretion
times, we can conclude that both planets undergo late water accretions with respect to their formation times.

As in Scenario 1, we also detect that several planetary embryos are temporarily captured in mean motion resonances with the Jupiter-mass planet.
In particular, in Simulation 6 of the Scenario 2, the planet which ends in the HZ is captured in the 2:1 mean motion resonance with the gas giant
during the evolution. This planet remains in this resonance for at least 1 Gyr.

\subsection{Scenario 3}
\label{sect:scenario3}

\begin{figure*}[htb!]
 \centering
\includegraphics[angle=0, width= 0.98\textwidth]{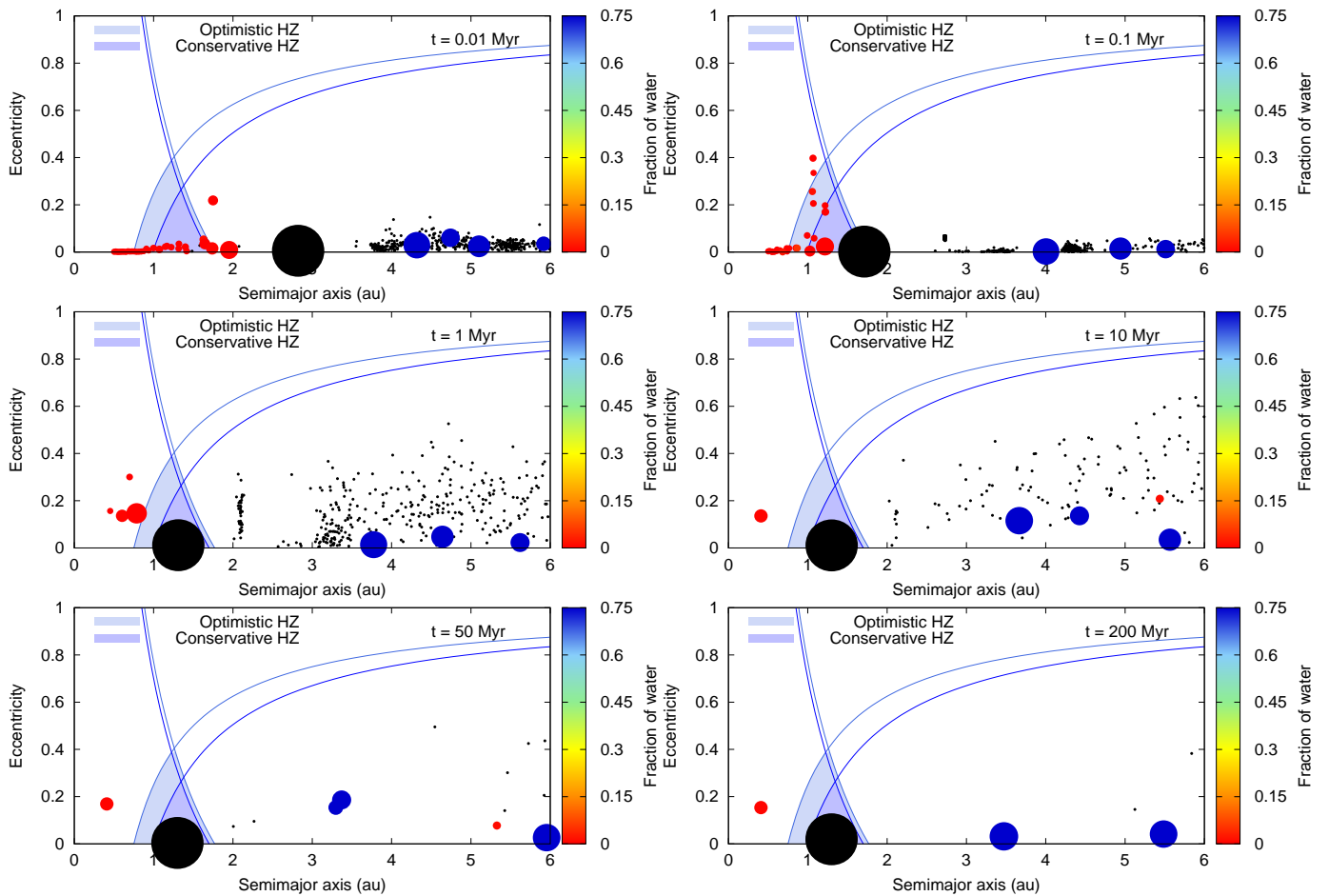}
 \caption{Snapshots of the dynamical evolution for Simulation 1 of Scenario 3. The colour code description is analogous to that of Fig. \ref{fig:time-evolution-Scen1-Sim16}.}
\label{fig:time-evolution-Scen3-Sim1}
\end{figure*}

We analysed the dynamical evolution of the systems in which the Jupiter-mass planet migrates up to $\sim 1.3$ au from the central star.
A main difference of this scenario with respect to the other two is that none of the simulations produce a planet in the HZ
of their systems after the 200 Myr of evolution. We consider the simulation labelled '1' as a representative of all the simulations
of this scenario. Figure \ref{fig:time-evolution-Scen3-Sim1} shows snapshots of the semi-major axis-eccentricity plane of Simulation 1,
as in the previous sections. The values of the time of each panel match with the times of Figs. \ref{fig:time-evolution-Scen1-Sim16}
and \ref{fig:time-evolution-Scen2-Sim15}.

From Fig. \ref{fig:time-evolution-Scen3-Sim1} we can see that the inner embryos are excited and reaches eccentricities $e \leq 0.4$ at $10^5$ yr.
However, the planetesimals show very low eccentricities with $e \leq 0.05$. At 1 Myr, most of the embryos are accreted by the
surviving ones ($\sim 2/3$ of the missing embryos), and the rest are ejected from the system, while the planetesimals acquire higher
eccentricities reaching $e \leq 0.6$. At 10 Myr there is only one planet internal to the giant, which remains there until the end of
the evolution, and four between the giant and 6 au from the central star. The number of planetesimals decreases significantly due to ejection. At 200 
Myr the planet inner to the giant remains in the same spot, while only two outer planets remain. We can also see that nearly all
planetesimals are removed from our studied region. In fact, $\sim$ 72\% of the initial planetesimal population is ejected from the system
while only 10\% remains with semi-major axes greater than 6 au.

As we see in the last panel of Fig. \ref{fig:time-evolution-Scen3-Sim1}, the giant is the only planet in the HZ. Because of its position
it is very difficult for another planet to survive inside the HZ during the evolution. To explain this, we followed the evolution of each
embryo and measured its distance with respect to the giant each time it enters in the HZ. Those results yield that, when a planet is
inside the HZ, the mutual distances to the giant can become as low as $\sim 0.1$ R$_{\text{Hill}}$, making the perturbation of the giant so large
that the planet cannot survive in the HZ.

\section{Discussion and conclusions}

In this paper, we studied the main aspects concerning the planet formation in systems with a migrating gas giant around solar-type stars.
First of all, we used a semianalytical model to determine the parameters of the protoplanetary disk able to form a Jupiter-mass
planet at the snow line. We proposed a late formation for the gaseous giant. In fact, when the giant is formed, the gaseous component
remained in the system for 1.5 $\times$ 10$^{5}$ yr, which represented the time that the Jupiter-mass planet has to migrate. From this,
we defined three scenarios of study, where the gas giant migrated up to intermediate distances of 2 au, 1.6 au, and 1.3 au with respect
to the central star. 
For each work scenario, the semi-analytic model also allowed us to determine the distributions of planetary embryos and planetesimals
until the Jupiter-mass planet was formed.
Finally, to analyse the main dynamical processes involved during and after the migration of the giant planet, we used
an N-body code. These simulations produced a wide variety of systems, forming planets with different masses, water contents,
and dynamical properties.

In particular, we were interested in studying the formation and evolution of planets in the HZ of the system. Our simulations produced
three different kinds of planets in the HZ. The first kind is represented by planets whose accretion seed
started the simulation beyond the snow line. Such planets ended up the simulation with significantly high water percentages
by mass (up to 75\% !) and owing to that, they were called water worlds.
The potential habitability of the water worlds is still under debate \citep{Abbot2012, Alibert2014, Kitzmann2015}. However, the water worlds would seem to be common in many planetary systems \citep{deElia2013, Ronco2014} and thus, we think that
they represent an interesting kind of exoplanets in the Universe. A detailed discussion about the potential habitability of water-rich planets with deep oceans can be found in \citet{Noack2016}.
The second kind of planets formed in the HZ is represented by those whose accretion seeds started the simulation
inside the snow line and their feeding zones were primarily associated to the inner region of the disk. Such planets were more
massive than the Earth and ended up the simulation with water percentages less than $\lesssim$ 1\% by mass. The planets
associated to this second kind are called Earth-like planets because they show features in their dynamical evolution
and final water percentages that result to be comparable to the Earth.
We think that such planets have an important astrobiological interest since they could achieve similar conditions to the Earth for the development and maintenance of life. Finally, the third kind of planets formed in the HZ is represented by those whose accretion seeds started the integration inside the snow line and did not accrete water-rich material along the evolution. These planets ended the simulation with no water and they are called dry worlds.

Those numerical simulations with a Jupiter-mass planet migrating to $\sim$2 au produced a total number of six out of twenty
planets in the HZ. In particular, two such planets were water worlds, with masses of 2.75 $M_{\oplus}$ and 3.57 $M_{\oplus}$ and final water contents of 75\% and 58.05\% by mass, respectively, which represent $\sim$7400 Earth oceans. Within this scenario of study, the other four planets formed in the HZ were Earth-like planets. In fact, such planets were super-Earths with masses ranging from 2.15 $M_{\oplus}$ to 3.85
$M_{\oplus}$, and final water contents less than 0.53\% by mass. Considering their masses and percentages of water, such planets showed
water contents less than 73 Earth oceans. It is worth noting that we analysed the long-term dynamical stability of the resulting systems,
inferring an efficient coexistence between a Jupiter-mass planet located at $\sim$2 au and terrestrial planets in the HZ with a considerably wide range of masses and water contents.

The numerical simulations with a Jupiter-mass planet migrating to $\sim$1.6 au formed a total number of three out of twenty planets
in the HZ. Such planets have masses ranging from 0.58 $M_{\oplus}$ to 3.51 $M_{\oplus}$ and final water contents less than 1.18\% by mass.
Considering their masses and percentages of water, such planets showed water contents less than 36 Earth oceans. It is worth emphasizing that
water worlds were not formed in this work scenario. Our analysis concerning the long-term dynamical stability of the resulting systems
suggested that a Jupiter-mass planet and an Earth-like planet coexisted inside the HZ. We conclude that this result is important
since it could have interesting implications in the search and discovery of exoplanets in the HZ around solar-mass stars that host a gas
giant.

There are two important differences between our results and those obtained by \citet{Fogg2009}. The first is related to
the mass of the planets surviving in the HZ of the systems of study. In general terms, the planets resulting from our simulations were
more massive than those produced by \citet{Fogg2009}. In fact, on the one hand, eight of ten planets surviving in the HZ in our
work had masses between 2.15 $M_\oplus$ and 3.85 $M_\oplus$. On the other hand, \citet{Fogg2009} formed a planet of
2 $M_\oplus$ in those scenarios where type I migration was excluded and migration of the giant planet was halted at distances close
to the outer limit of the HZ. The second relevant difference between our work and that from \citet{Fogg2009} is concerning
the final water content of the planets surviving in the HZ. The planet of 2 $M_\oplus$ produced by \citet{Fogg2009}
incorporated 40\% of its final mass from beyond the snowline at 2.7 au. According to these authors, one effect of giant planet migration
was to drive large quantities of icy material into the inner system, leading to the formation of water-rich terrestrial planets in the HZ.
Even though our numerical simulations formed two true water worlds in the HZ of the systems of study, six of the ten planets surviving in
the HZ showed final water contents less than 1.18\% by mass. Our work suggests that this kind of planet seems to be more common
than the water worlds in those scenarios where migration of the giant planet was halted close to the outer limit of the HZ.

The differences between our work and that carried out by \citet{Fogg2009} are due to a combination of multiple
factors. First, the mass of the protoplanetary disk used by \citet{Fogg2009} was a factor of 2.2 larger than that assumed in the
present work. In fact, even though both studies considered initial surface density profiles for gas and solids
taken from a MMSN model \citep{Hayashi1981}, this was scaled up in mass by factors of 1.35 and 3 for us and \citet{Fogg2009}, respectively. Moreover, these authors considered planetesimals of 10 km radius while a 0.5 $M_\text{Jup}$ giant
planet was assumed to migrate inward from 5 au. Unlike this, our work used planetesimals of 100 m radius and a Jupiter-mass planet
migrated inward from 3 au. Another significant difference between our work and that from \citet{Fogg2009} is related to the initial
distributions of embryos and planetesimals used in the N-body simulations. On the one hand, \citet{Fogg2009} assumed ad-hoc initial
masses for the population of planetary embryos of the system. In fact, the authors chose initial masses of 0.025 $M_\oplus$ and 0.1
$M_\oplus$ to represent embryos interior and exterior to the snowline, respectively. Moreover, the number of planetary embryos of each
N-body simulation was calculated by the authors assuming an average radial spacing between them of eight mutual Hill radii. Unlike this,
the work presented here used the semi-analytic model developed by \citet{Guilera2010} to analyse the formation of a giant planet
and the evolution of embryos and planetesimals during the gaseous phase. On the other hand, \citet{Fogg2009} assumed that $\sim$
90\% of the solid mass of the disk was contained in planetesimals. In our work, the semi-analytic model computed the evolution of the
planetesimal population during the gaseous phase. A more realistic treatment of the initial conditions associated to the populations
of embryos and planetesimals led to more robust results concerning the final masses and water contents of the terrestrial-like planets
that survived in the systems under study. Finally, it is important to remark that the N-body simulations developed by \citet{Fogg2009} were integrated for a time span of 30 Myr, while to test the stability of the resulting systems, a few simulations were extended
to 100 Myr. In the present work, we analysed the long-term dynamical stability of the systems integrating the simulations of
interest for a time span of 1 Gyr.

Finally, those numerical simulations with a Jupiter-mass planet migrating to $\sim$1.3 au did not form terrestrial planets in the HZ.
In fact, the gas giant efficiently removed planetary embryos and planetesimals from that region in all our N-body simulations.
This result is in agreement with that derived by \citet{Fogg2009}, who suggested that it is not conceivable that terrestrial planets survive in the HZ around a solar-type star if the gaseous giant's final orbit lies inside the region of $\sim$ 0.3 au - 1.5 au.

From the extrasolar systems discovered around solar-mass stars that host a Jupiter-mass planet between 1.5 au and 2 au, only two showed
properties comparable to the systems of our simulations. On the one hand, HD4208b is a gaseous giant with a minimum mass of
0.81 $\pm$ 0.03 Jupiter masses orbiting a central star of 0.88 $M_{\odot}$ with a semi-major axis of 1.65 $\pm$ 0.01 au and an
eccentricity of 0.05$^{+0.04}_{-0.03}$ \citep{Hollis2012}. On the other hand, HD82886b is a giant planet with a minimum
mass of 1.3 $\pm$ 0.1 Jupiter masses orbiting a central star of 1.06 $\pm$ 0.074 $M_{\odot}$ with a semi-major axis of 1.65 $\pm$ 0.06 au
and an eccentricity less than 0.27 \citep{Johnson2011}. Although a value of 0.27 for the eccentricity of HD82886b is
too large to consider it comparable to the systems of our simulations, \citet{Johnson2011} propose a possible solution which yield an eccentricity value of 0.066.
From our results, we considered that these two planetary systems could contain additional Earth-like planets in their HZs. 
We think that the PLATO (PLAnetary Transits and Oscillations of stars)\footnote{For more information see
\texttt{http://sci.esa.int/plato/}} mission will be able to confirm our results by detecting Earth-like planets in the HZs around Sun-like
stars that host giant planets at intermediate distances.

In the three work scenarios, many N-body simulations produced a system of super-Earths in the outer disk, more precisely,
between the location of the Jupiter-mass planet and 6 au. The gravitational microlensing technique is sensitive to planets
on wide orbits.
We think that the WFIRST (Wide-Field InfraRed Survey Telescope)\footnote{See \texttt{http://wfirst.gsfc.nasa.gov/}}
mission will help to detect this kind of planets since it will, in principle, be sensitive to all planets with mass $\gtrsim$
0.1 $M_{\oplus}$ and separations $\gtrsim$ 0.5 AU, including free-floating planets.

It is important to emphasise the limitations of the semi-analytic model and the N-body code used to carry out our study.
The semi-analytic model neglects the effect of gaseous envelopes for the embryos located in the inner and outer
regions respect to the position of the gas giant. However, the masses of the planetary embryos at the end of the gas phase
are less than 2.7 $M_{\oplus}$. Thus, we consider that such an approximation should not produce significant differences. Moreover,
the semi-analytic model does not take into account the processes of planetesimal fragmentation or pebble accretion. In the same way, we
did not include type I migration in our model since many quantitative aspects of such effect are still uncertain. As for the
planetesimal population, the semi-analytic model assumed only one species of planetesimals with a radius of 100 m. To consider
different sizes of planetesimals, we should have produced changes in the solid surface density profile at the end of the gaseous phase.
Our intention here is to describe the main simplifications of our semi-analytic model. However, we consider that such a model has allowed us
to carry out a good description of the evolution of the system during the gaseous phase and then, to derive more realistic conditions
for the planetary embryos and the planetesimal population. To analyse the sensitivity of the results to each of the effects mentioned
in the present paragraph is beyond the scope of this work.

The N-body MERCURY code also has some limitations that should be mentioned. On the one hand, collisions are treated as inelastic
mergers, conserving mass and water content. Thus, the mass and the final water content of the resulting planets represent upper limits.
\citet{Chambers2013} developed N-body simulations including fragmentation
and hit-and-run collisions rather than assuming that all impacts lead to a perfect merger of the colliding bodies. This improved model
was based on the results of hydrodynamical simulations of planetary impacts performed by \citet{Leinhardt2012} and \citet{Genda2012} that identified the boundaries of different collisional regimes and provided formulae for the mass
of the largest remnant. \citet{Chambers2013} carries out N-body simulations of terrestrial planet formation incorporating collisional fragmentation and hit-and-run collisions and then, he compares them to simulations in which all collisions were assumed to result in mergers. The final planetary systems formed in the two models were broadly similar. However, the author
found differences concerning the planetary masses and the time-averaged eccentricities of the final planets. Future N-body
simulations should include a more realistic treatment of the collisions in order to determine the orbital and physical characteristics
of the planets in more detail. On the other hand, \citet{Marcus2010} explores the effect of late giant impacts on
the final bulk abundance of water in super-Earths. According to the authors, if a giant impact between bodies of similar
composition occurs, the target planet either accretes materials in the same proportion, leaving the water fraction unchanged, or
loses material from the water mantle, decreasing the water fraction. \citet{Marcus2010} derives equations able to describe such processes. Other authors, such as \citet{Dvorak2015} quantify the fraction of water lost during an impact, inferring that during a collision a body might lose up to about 60$\%$ of its water content, depending on the energy and the geometry of the impact. Thus, future studies based on N-body simulations should include a more realistic model of the impacts in order to properly quantify the final water content.
We think that the observational evidence will allow us a refinement of our models, and therefore to strengthen our understanding
about the processes involved in the formation and evolution of planetary systems.

\begin{acknowledgements}
This work was partially financed by CONICET by grant PIP 0436/13. We thank the anonymous referee for valuable suggestions that helped
us improve the manuscript.
\end{acknowledgements}

\bibliographystyle{aa} % style aa.bst
\bibliography{Darriba_et_al_2017} % your references Yourfile.bib

\end{document}